# WORKING PAPER

# Improved Stock Price Forecasting Algorithm based on Feature-weighed Support Vector Regression by using Grey Correlation Degree

Quanxi Wang

# ABSTRACT

With the widespread engineering applications ranging from artificial intelligence and big data decision-making, originally a lot of tedious financial data processing, processing and analysis have become more and more convenient and effective. This paper aims to improve the accuracy of stock price forecasting. It improves the support vector machine regression algorithm by using grey correlation analysis (GCA) and improves the accuracy of stock prediction. This article first divides the factors affecting the stock price movement into behavioral factors and technical factors. The behavioral factors mainly include weather indicators and emotional indicators. The technical factors mainly include the daily closing data and the HS 300 Index, and then measure relation through the method of grey correlation analysis. The relationship between the stock price and its impact factors during the trading day, and this relationship is transformed into the characteristic weight of each impact factor. The weight of the impact factors of all trading days is weighted by the feature weight, and finally the support vector regression (SVR) is used. The forecast of the revised stock trading data was compared based on the forecast results of technical indicators (MSE, MAE, SCC, and DS) and unmodified transaction data, and it was found that the forecast results were significantly improved.

# 1. Introduction

With the application of computer science, data processing and data analysis of high-volume financial data have become easier and easier. People prefer to depend on computer to deal with financial problems rather than traditional artificial statistics, especial in stock price forecasting, one of the prevalent parts that many researchers focus on. The discuss with regard to the predictability of stock price in the actual market never ceases, since stock market is a complex nonlinear dynamic system and the estimation and extrapolation of classical nonlinear function values are far from being able to adapt to the complexity of the stock market.

Despite its complexity, many machine learning methods stand out to show their applications for stock price forecasting due to their exceptional nonlinear adaptability. Baba N and Kozaki M (1992) use a hybrid algorithm, combined the modified BP (back propagation) method with the random optimization method, to train the parameters in the neural network and develop the algorithm for forecasting stock prices in the Japanese market. Xing Chen (2001) and Weidong Meng and Taihua Yan present a method for stock market modeling and forecasting via fuzzy neural network based on T-S model. Yiwen Yang (2001) and Guizhong Liu and Zongping Zhang provide a method for predicting chaotic data with combining embedding theory and artificial neural networks and apply this method for stock market prediction. Tay F E H and Cao L (2007) propose a modified version of support vector machines to model non-stationary financial time series.

One of core problems of stock price forecasting is how to choose an effective quantitative financial technique. Support Vector Machine (SVM) is different from the traditional machine learning methods and has exceptional nonlinear adaptability. Its optimization goal is to minimize the confidence range value, and its optimization constraint is the training error. Besides that, The solution problem of SVM is finally transformed into the solution of quadratic programming problem, so the only global optimal solution can be obtained. We usually call SVM Regression SVR when SVM is applied for the regression problems. However, classical SVR method (c-SVR) has some limitations when applying for stock price forecasting since the prediction results are greatly affected by the input feature vector. Therefore, the feature weights of each feature vector should be considered in predicting the model establishment, and certain feature selection should be performed. Grey Correlation Analysis (GCA) is a method for measuring similarity among the factors based on nearness to the models which considering similarity and nearness respectively. By using this method, the feature weight of the feature vector can be determined. Interestingly, feature-weighted SVR (FWSVR) is not a brand-new model. In fact, it has already been examined by James N. K. Liu and Yanxing Hu (2012), developed a feature-weighted support vector machine regression algorithm for the China Shenzhen A-share market. However, until very recently FWSVR has not been considered as a mainstream SVR caused the limited test samples and untested robustness.

Another core problems of stock price forecasting is how to select precise forecasting factors. Stock price fluctuations are often affected by multiple factors. In the essence, stock prices fluctuate around their intrinsic value and are affected by the supply and demand of stocks according to the intrinsic value theory. At the same time , on the basis of behavioral finance theory, stock prices also fluctuate as investors' psychological expectations of various factors change. Hence, when considering the factors of stock price forecasting, we should combine both technical factors and behavioral factors.

The rest of the paper is structured as follows. In [Section 2](), we provide a detailed classification of the factors of stock price forecasting in both technical factors and behavioral factors. [Section 3]() covers two basic theories for stock price forecasting algorithm and the derivation of my improved algorithm named feature-weighed support vector regression by using grey correlation degree. [Section 4]() is devoted to the specific preparation for my improved algorithm. And two groups of forecasting results are presented in [Section 5]() and conclusive remarks are given in [Section 6]().

## 2. Impact factors of stock price forecasting

In this section, we mainly discuss the impact factors of stock price forecasting in two parts: technical impact factors and behavioral impact factors. And behavioral impact factors mainly includes weather indicators and emotional indicators.

### 2.1 Technical impact factors

Stock price forecasting research is inseparable from historical data, the most direct impact factors affecting the daily price of stocks is the performance of the market data of the day. They are the daily highest price $X_{11}$, the daily lowest price $X_{12}$, the daily trading volume $X_{13}$ and daily amount $X_{14}$.

Besides, according to the Prospect theory, traders are extremely sensitive to stock returns. Therefore, we include the daily close of the last day $X_{15}$ as stock returns' reference. Moreover, the performance of a single stock can be seen as a microcosm of the entire stock market trading and the market index is often introduced as a reference factor in forecasting. And in this article we use HS300 index $X_{16}$. In brief, we regard impact factors from $X_{11}$ to $X_{16}$ as technical impact factors.

### 2.2 Behavioral impact factors

The stock market trader's investment behavior is rooted in the investor's own cognition, emotion, attitude, and psychology, and is formed in the process of investor decision making and execution. Due to the complexity of stock market trader's psychology, there is not clear definition for the behavioral impact factors of stock market among the researchers around the world. In this article, we consider the behavioral impact factors of stock price as internal factors and external factors and we use weather factors to represent external factors and use emotional factors refer to as emotional factors.

### 2.2.1 Weather indicators

The weather effect of behavior means that a person's behavioral state can be affected by meteorology. Meteorological conditions are important factors that make up the human living environment. Meteorological conditions and their changes not only affect people's physical health, but also have a significant impact on people's psychological emotions.

The weather effect in stock market refers to the phenomenon of changes in the volatility of the stock market due to the weather. This paper mainly studies the impact of weather indicators on stock prices and selects temperature $X_{21}$, humidity $X_{22}$, atmospheric pressure $X_{23}$ and visibility $X_{24}$. In addition, this paper mainly collects urban weather indicators in Beijing and Shanghai, and uses the total amount of city transactions as the weight of the weather indicators.

### 2.2.2 Emotional indicators

At present, there are generally two methods for constructing emotional indicators for stock traders: direct investor emotional indicators like American Association of Individual Investor s' Index and indirect investor emotional indicators like turnover rate. Considering the objectiveness of indicators, this articles uses indirect investor emotional indicators. They are turnover rate in the past 20 trade days $X_{31}$, average idiosyncrasy rate (IVR) in the past 20 trade days $X_{32}$, AR $X_{33}$, ADTM $X_{34}$ and OBV $X_{35}$ and detailed computational procedure of each indicator are listed in appendix.

## 3. Basic theory of stock price forecasting and improved algorithm

In this section, we mainly discuss two basic theories for my stock price forecasting algorithm and my derivation of my improved algorithm named feature-weighed support vector regression by using grey correlation degree (FWSVR).

### 3.1 Support vector machine regression

The basic concept of support vector machine regression (SVR) is to map the training data set *x* in the input space *X* to the high-dimensional linear space *F* by a nonlinear mapping $\Phi$, thus converting the nonlinear function estimation problem in *X* into a linear function estimation problem in *F* and doing linear regression in space *F*. Finally return the regression result to the original input space *X* through inverse mapping, and obtain the regression result *y*. And the SVR function is formulated as follows.

$$f(x) = w\varphi(x) + b \tag{1}$$

where:

The regression estimated data set $\{x_i, y_i\}, (i = 1, 2, ..., N)$, $x_i \in R^n$ is system input vector and $y_i \in R$ is system output. $\varphi(x)$ is a nonlinear mapping from input space *X* to high-dimensional Helbert space. *w* is the linear combination of mapping function $\varphi(x)$ and it reflects the complexity of the data sample function. *b* is offset. The coefficients *w* and *b* could be estimated by minimizing:

$$R(f) = \frac{1}{2}\|w\|^2 + c\frac{1}{n}\sum_{i=1}^{n}L_\varepsilon(y_i, f(x_i)) \tag{2}$$

where:

$$L_\varepsilon(y, f(x)) = \begin{cases} 0, (|y - f(x)| \leq \varepsilon) \\ |y - f(x)| - \varepsilon, (|y - f(x)| > \varepsilon) \end{cases} \tag{3}$$

*C* is penalty factor. The larger the value of C, the greater the penalty for data that exceeds $L_\varepsilon$. $R(f)$ is composed of empirical error and a regularization term.

By applying Lagrange equation, dual problem conversion and kernel function $K(x_i, x_j) = \langle \varphi(x_i)\varphi(x_j) \rangle$, we could simplify the regression function as follows:

$$f(x) = \sum_{i=1}^{n}(\overline{\alpha_i} - \overline{\alpha_i}^*)K(x_i, x_j) + b \tag{4}$$

where:

$\overline{\alpha_i}^*$ is optimal solution. $\overline{\alpha_i} \neq \overline{\alpha_i}^*$

There are mainly two kinds of SVR models: $\varepsilon$-SVR and $\gamma$-SVR. This paper uses nonlinear $\varepsilon$-SVR.

**3.2 Grey correlation analysis**

The concept of grey correlation analysis (GCA) is originated from grey system theory by Julong Deng (1981). GCA measures the similarity according to the topological properties of the sequence curve in order to judge its similarity. The closer the curve topology distance is, the higher the correlation between the corresponding sequences, vice versa.

The algorithm step of GCA is as follows.

Step 1: Find the initial value of each sequence. Set system feature sequences $X_0 = \{X_0(1), X_0(2), \cdots, X_0(n)\}$. Then set the factor sequences $X_1, X_2, ..., X_m$. In this article, system feature sequences are the daily close price of stock and factor sequences are technical impact factors and behavioral impact factors $X_{11}, X_{12}, ..., X_{35}$.

Step 2: Calculate the range of sequences.

$$\Delta_i(k) = |X_0(k) - X_i(k)|, i = 1, 2, ..., m \tag{5}$$

Step 3: Calculate the maximum range of sequences and the minimum range of sequences.

$$M = \max_i \max_k \Delta_i(k), m = \min_i \min_k \Delta_i(k) \tag{6}$$

Step4: Calculate grey correlation coefficients.

$$\gamma_i(k) = \frac{m + \tau M}{\Delta_i(k) + \tau M}, \tau \in (0,1); k = 1, 2, ..., n; i = 1, 2, ..., m \tag{7}$$

where:

$\tau$ is resolution ratio, $\tau = 0.5$ in this paper.

Step 5: Calculate grey correlation degree.

$$\delta_i = \frac{1}{n} \sum_{k=1}^{n} \gamma_i(k), i = 1, 2, \cdots, m \tag{8}$$

$\delta_i$ could indicate the degree of correlation between each factor and the reference value (close price).

### 3.3 Feature-weighed support vector regression

In Euclid space, the distance of two vectors $X_i = (x_{i1}, x_{i2}, ..., x_{in}) \; X_j = (x_{j1}, x_{j2}, ..., x_{jn})$ could be represented as $d(X_i, X_j) = \sqrt{\sum_{k=1}^{n} |x_{ik} - x_{jk}|^2}$. And classical SVR also use this method to calculate the distance of sample points in high-dimensional space. And c-SVR is based on the assumption that all the features of the sample points supply the **same contribution** to the target output value. However, in real life, stock traders do not treat equally for the many impact factors of stock price, even the attitude towards the same impact factor varies from person to person. Hence, it is not reasonable to assume that all the features have the same weights. And grey correlation analysis exactly provides the degree of correlation between each factor sequence and the system feature sequence. So the contributions of each impact factors of stock price $w_i$ could be obtained by normalizing grey correlation degree $\delta_i$:

$$w_i = \frac{\delta_i}{\sum_{k=1}^{n} \delta_i(k)}, i = 1, 2, \cdots, m \tag{9}$$

Therefore, we could obtain the feature-weighed vector:

$$X_i = w_i(x_{i1}, x_{i2}, ..., x_{in}) \; X_j = w_j(x_{j1}, x_{j2}, ..., x_{jn})$$

Supposed *m* impact factors' weights are $W_i = (w_1, w_2, ..., w_m)$, so the distance of two feature vectors could be represented as:

$$d(X_i, X_j) = \sqrt{\sum_{k=1}^{n} w_k |x_{ik} - x_{jk}|^2} \tag{10}$$

The new kernel function is $K(W x_i, W x_j) = \langle \varphi(W x_i) \varphi(W x_j) \rangle$ and the final regression function of feature-weighed SVR (FWSVR) is formulated as follows:

$$f(x) = \sum_{i=1}^{n} (\overline{\alpha_i} - \overline{\alpha_i}^*) K(W x_i, W x_j) + b \qquad (11)$$

## 4. Preparation for the improved algorithm

In the last section, we conduct the basic regression function of FWSVR and the process to get its weights. In this section, we mainly discuss the basic preparation for FWSVR, including data preparation, data pre-processing, technical parameter setting and selection, evaluation system of forecast results setting and factor screening system.

### 4.1 Data preparation

Based on the previous studies on SVR for forecasting the China stock market, the range of the selected stock data for research only limit in several to dozens of stocks. So it could not completely proved the generalization of SVR algorithm in China stock market. However, considering that if the stocks of the entire market are selected as the research samples, the calculation speed is quite slower. Therefore, this paper selects the constituent stocks of the Shanghai and Shenzhen 300 Index (HS300) as the research object to test the performance of the algorithm in the entire market.

After the completion of the share reform in China's securities market, the stock index futures contract was officially listed and traded on April 16, 2009. On October 30, 2009, the China Growth Enterprise Market was opened for trading. With the gradual improvement of the system of China's securities market, the composition of investors in the market, the market environment and the original investment philosophy have changed accordingly. The investment style, investment philosophy and investment strategy of the market have begun to diversify. Small-cap stock investment strategy has become popular in stock market trading, and small-cap stock investment strategy is also the embodiment of stock trader's trading behavior adjustment. Hence, this paper selects the period when small-cap stock strategy is prevalent. The period range is from January 1, 2009 to December 31, 2016.

Classify the constituent stocks of the HS300 Index in 2009 based on the CITIC Level 1 industry classification. It shows that the constituent stocks of the HS300 cover 28 industries and have a certain representativeness.

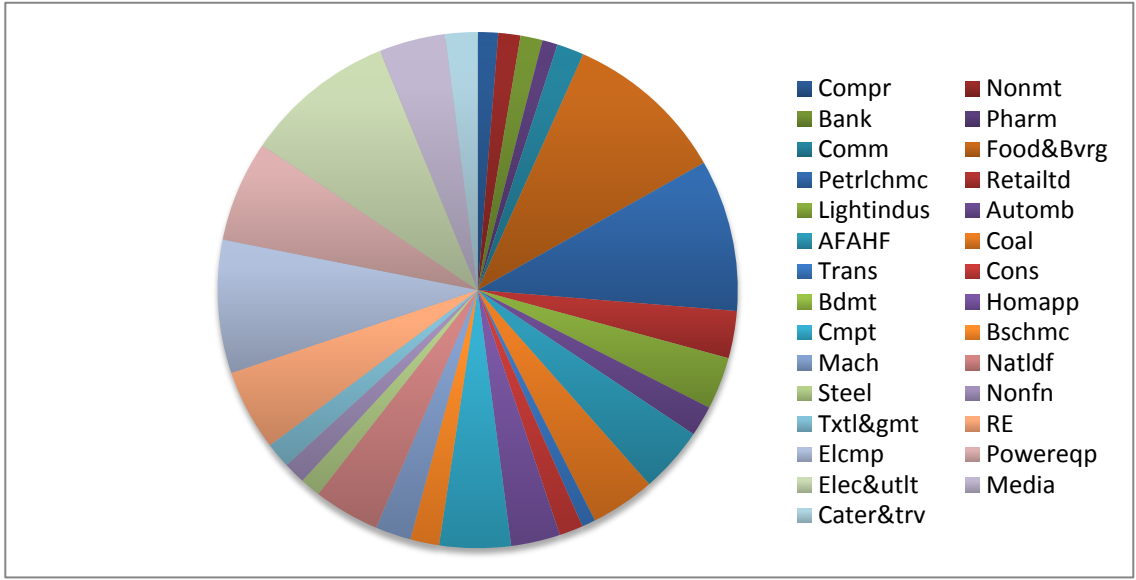

**Fig. 1** Industrial distribution of constituent stocks of HS300 index

All the technical impact factors and turnover rate and idiosyncrasy rate (IVR) of emotional indicators come from Wind Filesync database. AR, ADTM and OBV of emotional indicators come from uqer database. All the weather indicators come from NOAA（National Oceanic and Atmospheric Administration）database. Due to data compilation and sample issues, there are 298 stocks eventually included in the calculation.

**4.2 Data pre-processing**

Data pre-processing mainly consists of abnormal value handing and data regularization. First, the data for each stock during the suspension period is removed. Then use MAD method (Median Absolute Deviation) to screen and process extreme abnormal value in the sample. The MAD method is an improvement for the mean standard deviation method and its specific steps are as follows:

Step 1: Calculate sample standard deviation

$$md = median\{x_i\}, i = 1, 2, ..., n \qquad (12)$$

Step 2: Calculation threshold (median deviation)

$$MAD = median\{|x_i - md|\}, i = 1, 2, ..., n \qquad (13)$$

Step 3: Abnormal value correction. In this paper, 5 times the threshold is selected as the confidence interval.

$$if \ x_i > md + 5*MAD$$
$$x_i = md + 5*MAD$$
$$else \ if \ x_i < md - 5*MAD$$
$$x_i = md - 5*MAD \qquad (14)$$

After processing the corrected abnormal value, the sample data is regularized between intervals [-1, 1] using range normalization. On the one hand, it eliminates the influence of factors with large distribution range or small distribution range on the results, and on the other hand it can solve the over-fitting problem of support vector machine regression. It is formulated as follows:

$$x_i' = 2 \times \frac{x_i - \min_x}{\max_x - \min_x} - 1, i = 1, 2, ..., n \qquad (15)$$

where:

$x_i'$ is \transformed sample data, $x_i$ is raw sample data, $\min_x$ is the minimum of sample data and $\max_x$ is maximum of sample data.

**4.3 Technical parameter setting and selection**
The key to support vector machine regression prediction lies in the technical parameter setting and selection. It directly determines whether the optimal solution can be found, and determines the generalization ability and generalization of the model. This paper selects RBF (radial basis function) as kernel function in $\varepsilon$-SVR. In grey correlation degree calculation, the ratio between samples used to get the weight value and the total samples is 1:20. In SVR, the ratio between samples used to train and the samples used as testing data is 4:1. Moreover, this paper uses k-fold Cross Validation to obtain the other three core parameters: insensitive parameter $\varepsilon$, regularization parameter c and RBF kernel parameter $\gamma$. And its specific steps are as follows:

Step 1: Randomly divide the sample set S into k disjoint subsets, and the number of samples in each subset is m/k. These subsets are recorded separately *as $S_1, S_2,..., S_k$*.

Step 2:
   For each model $M_i$:
   *for j*=1 to *k*：
      Set $S_1 \cup ... \cup S_{j-1} \cup S_{j+1} \cup ... \cup S_k$ as training sets and train model $M_i$, then get the training function. Set $S_j$ as test sets and calculate the generalization error.

Step 3: Calculate the average generalization error for each model and select the parameters from the model $M_i$ that has the minimum average generalization error.

This paper uses Grid Regression Search to process the k-fold Cross Validation to obtain the optimal parameters.

**4.4 Evaluation system of forecast results**
In order to For measure the accuracy of forecast results, this paper mainly uses the following four evaluation indices:

Mean Squared Error (MSE) refers to the expected value of the square of the difference between the parameter estimate and the true value of the parameter. MSE could evaluate the degree of change in data. The smaller MSE values, the better precision the model is. It is formulated as follows:

$$MSE = \frac{1}{N} \sum_{t=1}^{N} (observed_t - predicted_t)^2 \qquad (16)$$

Mean Absolute Error (MAE) refers to the average of absolute error. MAE can better reflect the actual situation of the predicted value error. The smaller MAE values, the better precision the model is. It is formulated as follows:

$$MAE = \frac{1}{N}\sum_{t=1}^{N} |predicted_t - observed_t| \qquad (17)$$

Directional Symmetry (DS) indicates the correctness of the predicted direction of predicted value in terms of percentages. The bigger DS values, the better precision the model's trend forecasting ability is. It is formulated as follows:

$$DS = \frac{1}{N}\sum_{t=1}^{N} d_i * 100\%$$
$$d_i = \begin{cases} 1 & if\ (observed_t - observed_{t-1})(predicted_t - predicted_{t-1}) \geq 0 \\ 0 & otherwise \end{cases} \qquad (18)$$

Squared Correlation Coefficient (SCC) measures the degree of correlation between the parameter estimate and the true value of the parameter. The bigger SCC values, the better correlation the model is. It is formulated as follows:

$$SCC = \frac{(N\sum_{t=1}^{N} x_t * y_t - \sum_{t=1}^{N} x_t \sum_{t=1}^{N} y_t)^2}{(N\sum_{t=1}^{N} x_t^2 - \sum_{t=1}^{N} x_t \sum_{t=1}^{N} x_t)(N\sum_{t=1}^{N} y_t^2 - \sum_{t=1}^{N} y_t \sum_{t=1}^{N} y_t)} \qquad (19)$$

where:
$x$ is raw data, $y$ is predicting data.

### 4.5 Factor screening system

Because of the selection of 15 impact indicators including technical factors and behavioral factors, and not all of the selected impact indicators have a good ability to predict stock price. Therefore, a factor screening system should be set up before the final forecast to eliminate the indicators with poor predictive ability. In this way, we can not only avoid over-fitting of the final support vector machine regression, but also improve computational efficiency and reduces system resource waste. This paper mainly designs a two-step screening method based on grey correlation degree.

### 4.5.1 Preliminary screening based on grey correlation

Based on the explanation in Section 3, the grey correlation degree can better describe the similarity between the system reference sequence and the factor sequence. Hence, we take the first step to calculate the grey correlation degree of each forecasting indicator.

Step 1: Select stock $i$, $i=1,2,...,298$, calculate the grey correlation degree between all 15 impact indicators and the close price of stock $i$, and record them as $r_{i1}, r_{i2},..., r_{i15}$.

Step 2: Calculate the average grey correlation degree of all impact indicators.

$$\overline{r_i} = \frac{1}{n}\sum_{j=1}^{n} r_{ji}, n=298, i=1,2,...,15, \overline{r_i} \subset [0,1] \qquad (20)$$

Step 3: Set the threshold of average grey correlation degree as 0.6. The impact indicators with average grey correlation degree below 0.6 are marked as the indicators to be observed $h_1, h_2, ..., h_i, 0 \leq i \leq 15$ and these indicators need to follow the second step: random screening of SVR. The impact indicators with average grey correlation degree above 0.6 are marked as the basic forecasting indicators $l_1, l_2, ..., l_j, 0 \leq j \leq 15$ and these indicators could be directly used for FWSVR.

**4.5.2 Random screening of SVR**

The indicators to be observed $h_1, h_2, ..., h_i, 0 \leq i \leq 15$ need to process random screening of SVR. It refers to the selection of all basic forecasting indicators $l_1, l_2, ..., l_j, 0 \leq j \leq 15$ and a the indicator to be observed $h_i$ each time to form a quasi-predicted indicator group. Randomly select stocks with a total number of stocks of 1/10 as test samples and separately process c-SVR and FWSVR and compare average of the evaluation indices: *DS*, *MSE*, *MAE* and *SCC*. Count the number of evaluation indices that FWSVR perform better than c-SVR and record them as $N_{DS}$, $N_{MSE}$, $N_{MAE}$ and $N_{SCC}$. Repeat 3 times and then replace $h_i$. If the $N_{DS}$, $N_{MSE}$, $N_{MAE}$ and $N_{SCC}$ are lower than half number of the total samples, it indicates that the indicator $h_i$ has poor predictive ability .So eliminate the indicator $h_i$. If not, $h_i$ could be regarded as one of the basic forecasting indicators.

After the above two-step screening method, all the weather indicators and ADTM of emotional indicators are eliminated. All the technical factors and turnover rate, IVR, AR and OBV are allowed to the basic forecasting indicators.

## 5. Forecasting results

In this section, we mainly use basic forecasting indicators got from Section 4 as our final forecasting factors. The paper firstly uses only technical factors to exanimate whether FWSVR performs better than c-SVR. Then paper adds behavioral factors to test the effectiveness of behavioral and the advantages of FWSVR.

**5.1 Technical factors forecasting**

Since technical factors are widely used in the studies to forecast the stock price, the paper initially processes FWSVR and c-SVR separately by only using technical factors. Here is the figure 2 about normalizing grey correlation degree of technical factors for one stock. It can be seen that the weights of the five technical factors are relatively average, and all have a certain impact on the close price of the day. And the highest and lowest prices on the day performs more prominently.

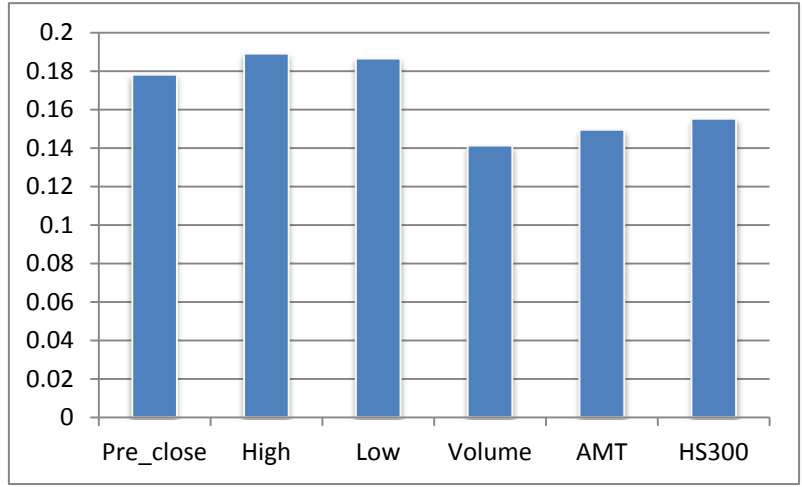

**Fig. 2** The weights of the five technical factors

Then use the *gridregression* function of toolbox *libsvm* in *python* 3.5 to process k-fold Cross Validation to obtain the optimal parameters in c-SVR: insensitive parameter $\varepsilon$, regularization parameter c and RBF kernel parameter $\gamma$. The default value of k is 10.

**Table 1** Several optimal parameters obtained from k-fold Cross Validation

|  | c | gama | epsilon |
|---|---|---|---|
| 000001.SZ | 64 | 0.003906 | 0.003906 |
| 000002.SZ | 64 | 0.03125 | 0.015625 |
| 000009.SZ | 64 | 0.007813 | 0.015625 |
| 000012.SZ | 64 | 0.007813 | 0.007813 |
| 000021.SZ | 64 | 0.015625 | 0.015625 |
| 000024.SZ | 64 | 0.03125 | 0.03125 |
| 000027.SZ | 64 | 0.003906 | 0.03125 |
| 000031.SZ | 64 | 0.007813 | 0.003906 |
| 000039.SZ | 64 | 0.015625 | 0.007813 |

Finally, based on the optimal parameters, process FWSVR and c-SVR separately. Figure 3 is a comparison of the raw data and predicted results of two stocks of 000001 and 600143 under FWSVR.

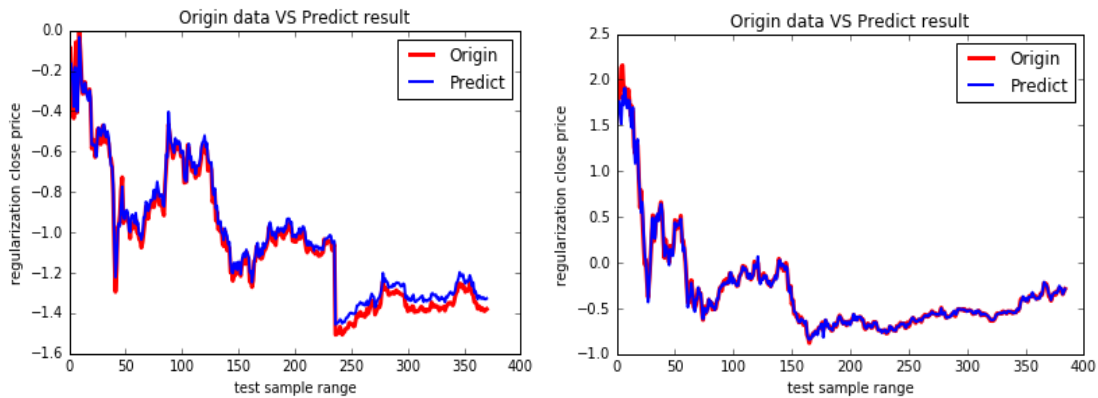

(a) 000001 raw data and predicted results    (b) 600143 raw data and predicted results

**Fig. 3** Comparison of the raw data and predicted results under FWSVR

Count the number of FWSVR performing better than C-SVR under each evaluation index.

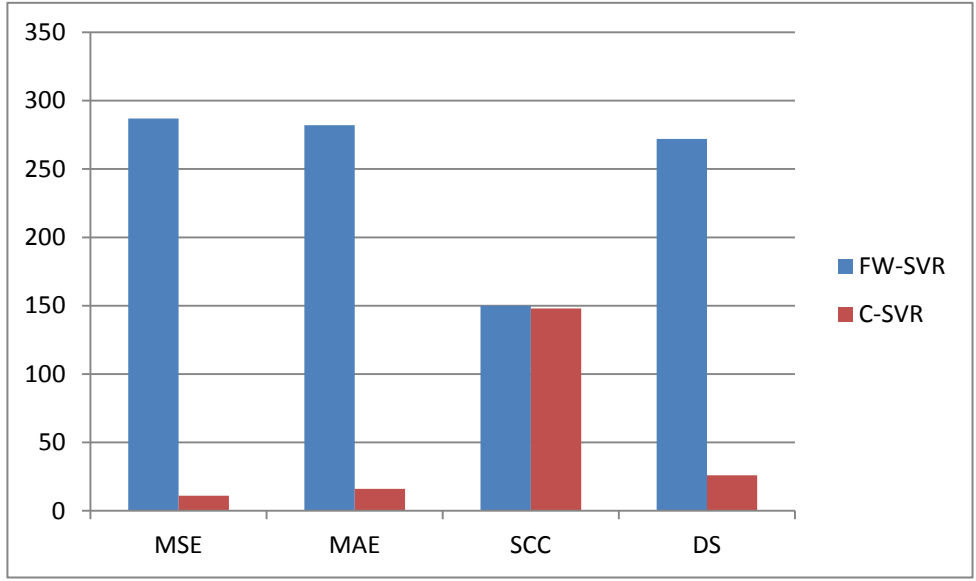

**Fig. 4** The number of FWSVR performing better than C-SVR under each evaluation index

The results show:

Except for the SCC index which FWSVR is slightly better than the C-SVR, the other three evaluation indices of FWSVR are obviously dominant.

By comparing the SCC index between FWSVR and C-SVR, it indicates that the difference between both SCC indices is lower than 0.01, which we could conjure that FWSVR has less improvement to SCC index.

Further statistics to analyze the improved evaluation indices, MSE has the highest improvement with an average increase of 65.30%. The improvement of MAE is second, and the average improvement is 37.02%. DS is also improved with an average increase of 23.52%. SCC has the lowest improvement with an increase of 5.38%.

Therefore, the results on the one hand indicate that the technical factors can indeed predict the stock. On the other hand, it has also proved that FWSVR model can improve the C-SVR model and improve the error and trend.

## 5.2 Technical factors and behavioral factors forecasting

### 5.2.1 Comparison of forecasting results to results in C-SVR

Add behavioral factors (turnover rate, IVR, AR and OBV) to the basic forecasting indicators.

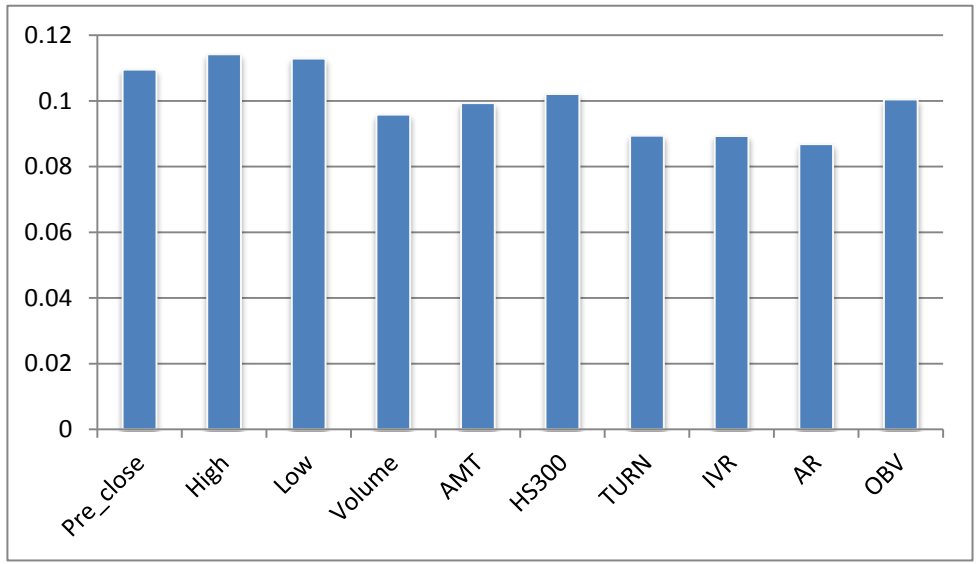

**Fig. 5** The weights of the 10 impact factors

The figure about normalizing grey correlation degree of each factor for one stock is shown above. It can be seen that the weights of the 10 factors are alsorelatively average, and all have a certain impact on the close price of the day. And the highest and lowest prices on the day still performs more prominently.

After obtaining the optimal parameters from k-fold Cross Validation, process FWSVR and c-SVR separately. Figure **5** is a comparison of the raw data and predicted results of two stocks of 000001 and 600143 under FW-SVR.

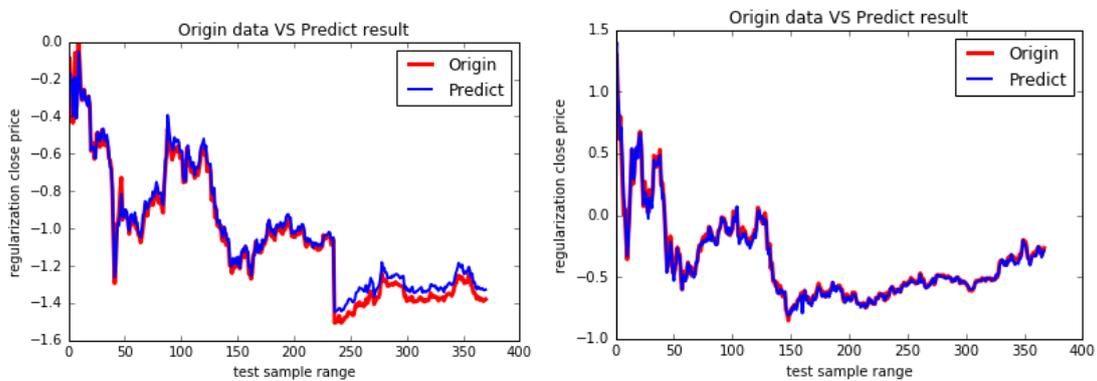

(a) 000001 raw data and predicted results    (b) 600143 raw data and predicted results

**Fig. 6** New comparison of the raw data and predicted results under FWSVR

Count the number of FWSVR performing better than C-SVR under each evaluation index.

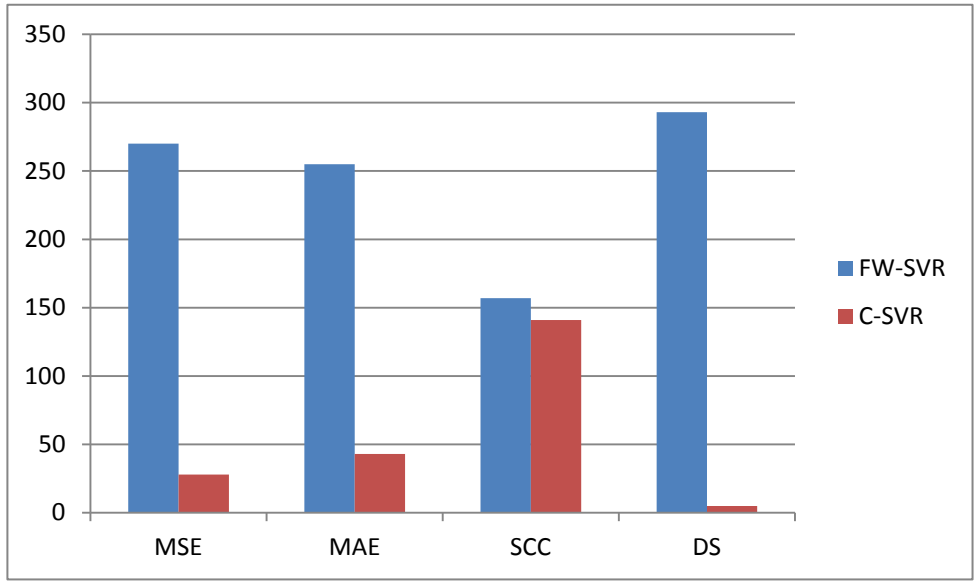

**Fig. 7** The new number of FWSVR performing better than C-SVR under each evaluation index

The results show:

After adding behavioral factors, all the evaluation indices of FWSVR are obviously dominant in comparison to C-SVR It indicates that impact factors do not change the stability of FWSVR model and FWSVR model has a strong robustness.

Further statistics to analyze the improved evaluation indices, MSE has the highest improvement with an average increase of 66.59%. The improvement of MAE is second, and the average improvement is 46.56%. DS is also improved with an average increase of 38.26%. SCC has the lowest improvement with an increase of 3.49%.

Therefore, the results on the one hand indicate that the technical factors and behavioral factors can indeed predict the stock. On the other hand, it has also proved that FWSVR model can improve the C-SVR model and improve the error and trend.

**5.2.2 Comparison of forecasting results to that only contain technical factors**

From the number of stocks in which FWSVR is better than C-SVR under each evaluation index:

**Table 2** The number of stocks in which FWSVR is better than C-SVR for each evaluation index

|         | MSE | MAE | SCC | DS  |
|---------|-----|-----|-----|-----|
| FW-SVR1 | 287 | 282 | 150 | 272 |
| FW-SVR2 | 270 | 255 | 157 | 293 |

where:
FW-SVR1 refers to the number of stocks that only include technical factors, and FW-SVR2 refers to the number of stocks that include both technical and behavioral factors.

The results show that after considering the behavioral factors, FWSVR model reduces the number of prediction errors and improves the trend and correlation of prediction results.

From the average degree of improvement of evaluation indicators:

Table 3 The average degree of improvement of evaluation indicators for each evaluation index

|  | MSE | MAE | SCC | DS |
|---|---|---|---|---|
| FW-SVR1 | 65.30% | 37.02% | 5.38% | 23.52% |
| FW-SVR2 | 66.59% | 46.56% | 3.49% | 38.26% |

The results show that after considering the behavioral factors, FWSVR model has a certain degree of improvement on the prediction error and prediction trend, and the degree of improvement of the prediction correlation has a certain degree of reduction.

From the perspective of each stock:

Separately count evaluation indices of each stock that with only technical factors and including both technical and behavioral factors. For each stock, if the latter evaluation index is better than the former, then count as 1 for corresponding factors.

Table 4 The number of stock which the latter evaluation index is better than the former

|  | MSE | MAE | SCC | DS |
|---|---|---|---|---|
| FW-SVR3 | 257 | 256 | 31 | 26 |
| FW-SVR4 | 41 | 42 | 267 | 272 |

The results show that adding behavioral factors can significantly improve the trend and correlation of prediction results for each stock, but at the same time it will improve the prediction error.

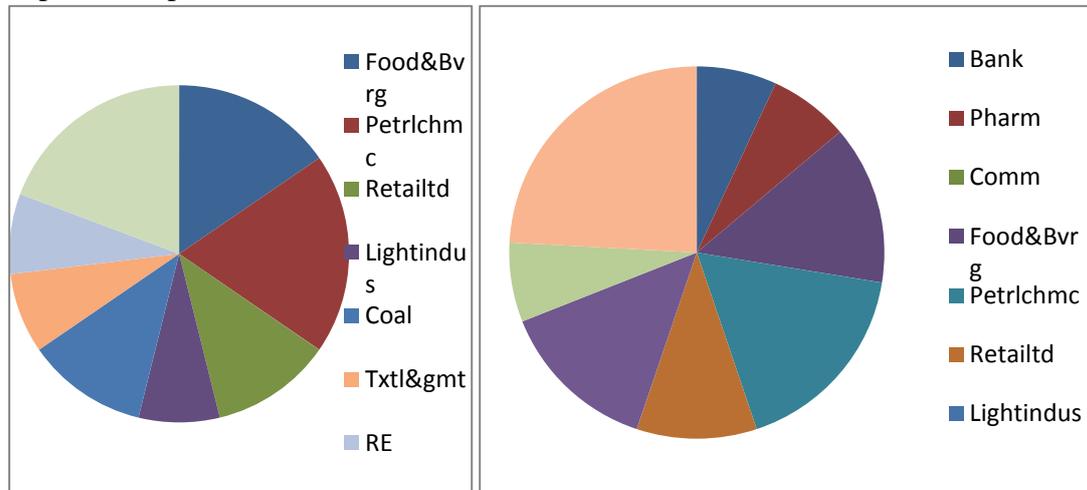

(a) SCC industrial distribution of improved stocks (b) DS industrial distribution of improved stocks

Fig. 8 Industrial distribution of improved stocks

Further statistics on the industry classification of stocks with improved evaluation indicators (CITIC Level 1). By calculating the proportion of the industry, it was found that behavioral factors are more prominent in the power equipment, textile and apparel, coal, commercial retail, petrochemical and food and beverage industries, especially in the power equipment and petrochemical industries that the improved stocks dominate large proportion.

In summary, after adding behavioral factors, the improvement of evaluation indices has the characteristics of small amount and great improvement, and the trend of prediction results has been significantly improved, but the number of stocks with improved prediction error is small. Moreover, from the perspective of a single stock, the improved stocks are mainly concentrated in power equipment, petrochemicals, textiles and clothing, coal, commercial retail and food and beverage, reflecting to some extent the behavioral factors have certain predictive power in these industries.

**5.3 Ideas for real trading strategy**

Based on the aforementioned forecasting results, it is demonstrated that FWSVR model has a certain effect on the stock to improve the accuracy of stock forecasting. As a result, this model can be used as a stock selection system in stock trading strategies and this paper provides some ideas for applying this method to real trading. And the specific steps of the stock selection strategy constructed are as follows:

Step 1: Set stock pools (the range of stocks), stock selection time span and factor pools. For Chinese stock market, generally we can choose HS 300 constituent stocks or CSI 500 constituent stocks. In order to ensure the accuracy of stock selection, the stock selection time span should be more than 3 years. Factor pools should include behavioral factors based on the premise that technical factors are included.

Step 2: Run the FWSVR model program for data back testing. Through the data processing of the program, the setting and selection of technical parameters, the factor screening and other steps, the result values of each evaluation index of each stock under the time span can be obtained.

Step 3: The stocks are screened according to the performance of the stock evaluation indices, and the stocks with better evaluation indicators are included in the forecast set. Run the FWSVR model program for the next-day stock price forecasting in the forecast set. If a positive rate of return is generated, the stock can be bought and vice versa. Also, perform daily data updates for stocks in the forecast set and repeat the steps from Steps 2 to Steps 3.

## 6. Conclusions

In this paper, we expand the range of impact factors for feature-weighted SVR algorithm and set a longer period to exanimate it robustness. We first divide the factors affecting the stock price movement into behavioral factors and technical factors and the behavioral factors mainly include weather indicators and emotional indicators. And then measure relation through the method of grey correlation analysis. The relationship between the stock price and its impact factors during the trading day, and this relationship is transformed into the characteristic weight of each impact factor. The weight of the impact factors of all trading days is weighted by the feature weight, and finally the support vector regression (SVR) is used.

In comparison to classical SVR model, forecasting results are significantly improved in FWSVR model and FWSVR has robustness and generality. In addition, we compare the two situations that contain only technical factors and that include both technical factors and behavioral factors. The results indicate that both technical factors and behavioral factors can indeed be applied to predict the stock. Technical factors are more applicable, and the application of behavioral factors is narrower. But the behavioral factors also have a greater improvement on the average of the evaluation indices, especially in power equipment, petrochemicals, textiles and clothing, coal, commercial retail and food and beverage industries.

# Appendix

The calculation of emotional indicators:

idiosyncrasy rate (IVR) $X_{32}$ refers to market-independent non-system volatility from Fama-French three-factor model. It is formulated as follows:

$$r_i - r_f = \beta_1(r_m - r_f) + \beta_2 HML + \beta_3 SMB + IVR_t \tag{A.1}$$

where:
Factor *HML* represents the high market-to-equity ratio of stock portfolio yields minus the low-to-market ratio stock portfolio yields, and factor *SMB* represents small-cap stocks yields minus large-cap stock returns.

AR $X_{33}$ reflects market popularity. It is based on the opening price of the day, by comparing the highest and lowest prices of the day with the market price of the day, to reflect the position of the opening price in the stock price through a certain period of time. It is formulated as follows:

$$AR = \frac{\sum_{i=1}^{N}(P_{max,i} - P_{start,i})}{\sum_{i=1}^{N}(P_{start,i} - P_{min,i})} \tag{A.2}$$

where:

$P_{max,i}$ is the highest price for the i-day, $P_{min,i}$ is the lowest price for the i-day, $P_{start,i}$ is the opening price for the i-day, $N= 26$.

ADTM $X_{34}$ refers to the difference between the upward fluctuation range of the opening price and the downward fluctuation range to describe the high and low. Its steps are as follows:

Step 1: *if* opening price<=opening price of last day，DTM=0；

　　　　else $DTM = \max((P_{max} - P_{start}), (P_{start} - P_{last}))$

Step 2: *if* opening price>=opening price of last day，DBM=0；

　　　　*else* $DBM = \max((P_{start} - P_{min}), (P_{start} - P_{last}))$

Step 3: $STM = \sum_{i=1}^{N} DTM_i$, $SBM = \sum_{i=1}^{N} DBM_i$

Step 4: *if* STM>SBM,ADTM=(STM-SBM)/STM；

　　　　*else if* STM<SBM,ADTM=(STM-SBM)/SBM；

　　　　*else* ADTM=0

where:

$P_{last}$ is opening price of last day, $N=23$.

OBV $X_{35}$ refers to the change in the trading volume of the stock market to measure the driving force of the stock market, so as to judge the trend of the stock price. It is formulated as follows:

$$OBV_i = OBV_{i-1} + \text{sgn}*Vol_i$$

$$\text{sgn} = \begin{cases} 1 & P_{start,i} \geq P_{last,i} \\ -1 & P_{start,i} < P_{last,i} \end{cases} \quad (A.3)$$

where:

$Vol_i$ is the number of stock transactions for the i-day, the initial value of OBV is generally replaced by the volume of the first day.